\UseRawInputEncoding
\documentclass[reprint]{revtex4-2}
\usepackage{amsmath}
\usepackage{subfig}
\usepackage{graphicx}
\usepackage{epstopdf}
\usepackage{verbatim}
\usepackage{color}
\usepackage{ragged2e}
\usepackage{ulem}
\usepackage{amssymb,amsfonts}
\usepackage{latexsym}
\usepackage{epsfig}
\usepackage{amssymb}
\usepackage{color}
\usepackage{bm}
\usepackage{bbm}
\usepackage{lipsum,mwe}
\usepackage{soul}
\newcommand{\be}{\begin{equation}}
\newcommand{\ee}{\end{equation}}

\newcommand{\bea}{\begin{eqnarray}}
\newcommand{\eea}{\end{eqnarray}}

\begin{document}
 %\title{\textcolor{red}{The critical end point of chiral phase transition meets the critical end point of nuclear liquid-gas phase transition}}
 %\title{\textcolor{red}{The chiral phase transition meets the nuclear liquid-gas phase transition}}
 %\title{The negative kurtosis $\kappa\sigma^2$ of baryon number fluctuation from nuclear liquid-gas phase transition}
 \title{The baryon number fluctuation $\kappa\sigma^2$ as a probe of nuclear matter phase transition at high baryon density}
\author{Kun Xu $^{1}$}
\thanks{xukun21@ucas.ac.cn}

\author{Mei Huang $^{1}$ }
\thanks{huangmei@ucas.ac.cn}

\affiliation{$^{1}$ School of Nuclear Science and Technology, University of Chinese Academy of Sciences, Beijing 100049, China}

\begin{abstract}
Two critical end points (CEPs) of the chiral phase transition and the nuclear liquid-gas phase transition show up at finite baryon chemical potential. The kurtosis $\kappa\sigma^2$ of baryon number fluctuation on the $T-\mu_B$ plane is positive  on the first-order side and negative on the crossover side along the phase boundary. The freeze-out line extracted from the heavy ion collisions crosses between these two phase boundaries, one can observe a peak of $\kappa\sigma^2$ around the collision energy $5 {\rm GeV}$ near the CEP of the chiral phase transition, and negative $\kappa\sigma^2$ at low collision energies due to the CEP of the nuclear liquid-gas phase transition. This expalains the experimental measurement of $\kappa\sigma^2$ at the collision energies of 2.4 GeV at HADES and 3 GeV and 7.7-200 GeV at STAR for most central collision.  Thus we propose that the baryon number fluctuation $\kappa\sigma^2$ can be used as a probe of nuclear matter phase structure at high baryon density.
\end{abstract}
\pacs{Null }
\maketitle
%forming a cliff? forming a canyon?
\textit{Introduction:} Exploring the QCD phase diagram especially at high bayron densities has been one of the main goal of high energy nuclear physics, which is essential to understand the inner structure of neutron stars(NS), gravitational waves emitted from NS mergers, and the formation and evolution of pulsars. It has been predicted from effective QCD models that a critical end point(CEP) exists at finite baryon chemical potential \cite{Pisarski:1983ms,Stephanov:1998dy,Hatta:2002sj,Stephanov:1999zu,Hatta:2003wn,Schwarz:1999dj,Zhuang:2000ub}, and the search of the CEP through heavy ion collisions has become one of the most important goals at high baryon densities with heavy ion collisions, for example, the beam energy scan (BES) program at relativistic heavy ion collision (RHIC) at BNL \cite{STAR:2010mib,STAR:2010vob,STAR:2013gus,Luo:2017faz,STAR:2020tga,STAR:2022etb}, as well as at future facilities like FAIR at Darmstadt, NICA in Dubna and HIAF in Huizhou.
%\cite{STAR:2005gfr,BRAHMS:2004adc,PHOBOS:2004zne,PHENIX:2004vcz}, 

QCD phase structure at high baryon density has not been well understood, due to the sign problem in lattice QCD calculations at finite baryon chemical potential, and the gluon degrees of freedom can not be effectively taken into account at finite baryon density. It was proposed in \cite{Hatsuda:2006ps} that there is an extra CEP with a crossover hadron-quark transition at zero temperature and high baryon density, which is important to determine the neutron star structure \cite{Huang:2022mqp} as well as the neutron stars merging. Also  the quarkyonic phase was proposed in the moderate baryon density region in \cite{McLerran:2007qj}.

%\cite{Klevansky:1992qe,Schaefer:2007pw,Fu:2019hdw} 
The fluctuation of conserved charge, e.g., the net baryon number, net strangeness and net electric charge, can be a probe of the CEP \cite{Hatta:2002sj,Stephanov:1999zu,Hatta:2003wn,Stephanov:2011pb}. RHIC's first phase of Beam Energy Scan (BES-I) has shown a non-monotonic behavior of baryon number fluctuation $C_{4}/C_{2}$, i.e., the kurtosis $\kappa\sigma^2$ \cite{Luo:2017faz} in the range of collision energy $\sqrt{s}=7.7 \sim 200 {\rm GeV}$ \cite{STAR:2020tga}. This non-monotonic behavior of baryon number fluctuation can be described in effective chiral models \cite{Fu:2016tey,Stephanov:2011pb,Fan:2016ovc,Li:2018ygx}. It was pointed out in \cite{Li:2018ygx} that the location of the CEP and the relative location of the freeze-out line and phase boundary are essential to determine the behavior of $\kappa\sigma^2$ along the collision energy $\sqrt{s}$, and a peak structure at $\sqrt{s}\sim 5{\rm GeV}$ related to the CEP of chiral phase transition was predicted. Recently, results of the kurtosis $\kappa\sigma^2$ for most central collision from RHIC at 3GeV \cite{STAR:2022etb} and from HADES at 2.4GeV \cite{HADES:2020wpc} shows negative value, which are lack of theoretical explanations. We will offer an understanding from theoretical side.

Models including quarks like Nambu-Jona--Lasinio model exhibit chiral phase transition at finite temperature and finite baryon chemical potential \cite{Klevansky:1992qe}, and the corresponding CEP of chiral phase transition, ${\rm CEP}_{\chi}$, could affect $C_4/C_2$\cite{Stephanov:2011pb}. However, the properties of hadronic phase are ignored in quark models. It has been well-known that, except the CEP for chiral phase transition, nuclear Liquid-Gas(LG) phase transition also exhibts a CEP, ${\rm CEP}_{\rm LG}$, at large baryon chemical potential and near zero temperature, \cite{Vovchenko:2016rkn,Glendenning:1992vb,Shao:2017yzv,He:2022yrk}. It is reasonable to consider hadrons at low collision energy where low temperature and large baryon chemical potential could be reached. The universal feature of CEP for either chiral phase transition or Liquid-Gas phase transition is the same \cite{{Stephanov:2011pb}}, specifically, $\kappa\sigma^2$ is negative in the crossover side along the phase transition line with temperature larger than $T_{{\rm CEP}}$.

In this Letter, we investigate the baryon number fluctuation $\kappa\sigma^2$ as a function of collision energy in a hybrid model including both ${\rm CEP}_{\chi}$ and ${\rm CEP}_{\rm LG}$. Numerical calculation shows $\kappa\sigma^2$ along the freeze-out line are well in agreement with experiment data. Considering that models including baryons and quarks are out of reach, instead we take into their contributions via Polyakov-Loop Nambu-Jona--Lasinio model (PNJL) and Walecka model separately, and connected by Gibbs condition, i.e., the dominant degree is determined by the grand potential at given temperature and baryon chemical potential, if $\Omega_{B}>\Omega_{q}$, PNJL model is used otherwise Walecka model, i.e., $\Omega=\text{min}\{\Omega_B,\Omega_q\}$.

\textit{The Polyakov-Loop Nambu-Jona--Lasinio model:} In the realistic PNJL model\cite{Bhattacharyya:2016jsn}, quarks are described by 3-flavor which takes into account 8-quark interactions, the grand potential under mean field approximation(MFA) is given below:
\begin{eqnarray}
& &\Omega_{{\rm PNJL}} =U(\Phi,\bar{\Phi},T)+\nonumber \\
  & &g_s\sum_{f}{\sigma_f^2}-\frac{g_D}{2}\sigma_u\sigma_d\sigma_s
        +3\frac{g_1}{2}(\sum_{f}{\sigma_f^2})^2+3g_2\sum_f{\sigma_f^4} \nonumber \\
        & &-6\sum_{f}\int_{-\Lambda}^\Lambda \frac{d^3p}{(2\pi)^3} E_f  -2T\sum_{f}\int_{-\infty}^{\infty} \frac{d^3p}{(2\pi)^3} \times \biggl\{\nonumber \\
& &  \ln\biggl[1+3\Phi e^{-\frac{E_f-\mu_f}{T}}+3\bar{\Phi}e^{-2\frac{E_f-\mu_f}{T}}+e^{-3\frac{E_f-\mu_f}{T}}\biggl] +\nonumber \\
& &  \ln\biggl[1+3\bar{\Phi} e^{-\frac{E_f+\mu_f}{T}}+3\Phi e^{-2\frac{E_f+\mu_f}{T}}+e^{-3\frac{E_f+\mu_f}{T}}\biggl] \biggl\}, 
\end{eqnarray}
where $\sigma_f$ is the quark condensates and $f$ takes $u,d$ for two light flavors while $s$ for strange quark.  $E_f=\sqrt{p^2+M_f^2}$ with $M_f$ the dynamically generated constituent quark mass $M_f=m_f-2g_s\sigma_f+\frac{g_D}{4}\sigma_{f+1}\sigma_{f+2}-2g_1\sigma_f(\sum_{f'}{\sigma_{f'}^2})-4g_2\sigma_f^3$. If $\sigma_f=\sigma_u$, then $\sigma_{f+1}=\sigma_d$ and $\sigma_{f+2}=\sigma_s$, and so on in a clockwise manner. Considering NJL model is non-renormalised, a cutoff in the vacuum integration is applied while no constrains to thermal terms. And there is no gluon in the NJL model, thus, Polyakov loop $\Phi$/$\bar{\Phi}$ are considered to take the confinement into account effectively, while the potential $U(\Phi,\bar{\Phi},T)$ is included to mimic the gluon self-interaction which reads \cite{Ghosh:2007wy}:
\begin{equation}
    \frac{U}{T^4}=-\frac{b_2(T)}{2}\bar{\Phi}\Phi-\frac{b_3}{6}(\Phi^3+\bar{\Phi}^3)+\frac{b_4}{4}(\Phi\bar{\Phi})^2,
\end{equation}
$b_2(T)$ is a temperature dependent coefficient which is chosen to have the form of $b_2(T)=a_0+a_1 \frac{T_0}{T}\exp(-a_2 \frac{T}{T_0})$. The parameters in NJL part are fixed by vacuum properties and take $\Lambda=637.72\text{MeV}$, $m_{u,d}=5.5\text{MeV}$, $m_{s}=183.468\text{MeV}$, $g_s \Lambda^2=2.914$, $g_D \Lambda^5=75.968$, $g_1 =2.193\times 10^{-21}\text{MeV}^{-8}$, $g_2 =-5.89\times 10^{-22}\text{MeV}^{-8}$, while the parameters of Polyakov loop part are fixed by global fitting of the pressure density to Lattice data at zero chemical potential which reads $T_0=175$MeV, $a_0=6.75$, $a_1=-9.8$, $a_2=0.26$, $b_3=0.805$, $b_4=7.555$.

For quark matter in equilibrium state, the quantities can be solved by four gap equations:
\begin{equation}
\frac{\partial \Omega_{{\rm PNJL}}}{\partial \sigma_{l/s}}=0,    \frac{\partial \Omega_{{\rm PNJL}}}{\partial \Phi/\bar{\Phi}}=0,
\end{equation}
with the solutions $\sigma_{l/s}$ and $ \Phi/\bar{\Phi}$, the pressure at given temperature and baryon chemical potential$\mu_B$ can be determined via $P_q(T,\mu_B)=P_{q,0}(T,\mu_B)-P_{q,0}(0,0)$ with $P_{q,0}(T,\mu_B)=-\Omega_{{\rm PNJL}}(T,\mu_B)$. 

\textit{Walecka model:} For hadronic phase, we only consider the contributions of nucleons(N)($n$ and $p$) and hyperons(H)($\Lambda$ and $\Xi$), as well as the scalar mesons $\sigma$, $\xi$ and vector mesons  $\omega$ and $\phi$, and the interactions among them are described by Walecka model, the Lagrangian of which is \cite{He:2022yrk,Bunta:2004ej}
\begin{eqnarray}
    &&\mathcal{L}_{\text{Wal}}=\sum_{B}\bar{\psi}_B\biggl[\gamma^{\mu}(i\partial_{\mu}-g_{\omega B}\omega_{\mu}-g_{\phi B}\phi_{\mu})-\nonumber\\
    &&(M_B-g_{\sigma B}\sigma-g_{\xi B}\xi)\biggr]\psi_B+\frac{1}{2}(\partial_{\mu}\xi\partial^{\mu}\xi-m_{\xi}^2\xi^2)+\nonumber\\
    &&\frac{1}{2}(\partial_{\mu}\sigma\partial^{\mu}\sigma-m_{\sigma}^2\sigma^2)-\frac{1}{3}b_{\sigma}M_N(g_{\sigma}\sigma)^3-\frac{1}{4}c_{\sigma}(g_{\sigma}\sigma)^4+\nonumber\\
    &&\frac{1}{2}m_{\omega}^2\omega_{\mu}\omega^{\mu}-\frac{1}{4}\omega_{\mu\nu}\omega^{\mu\nu}+\frac{1}{2}m_{\phi}^2\phi_{\mu}\phi^{\mu}-\frac{1}{4}\phi_{\mu\nu}\phi^{\mu\nu},
\end{eqnarray}
where $\omega_{\mu\nu}=\partial_{\mu}\omega_{\nu}-\partial_{\nu}\omega_{\mu}$, $\phi_{\mu\nu}=\partial_{\mu}\phi_{\nu}-\partial_{\nu}\phi_{\mu}$. Here we use MFA, i.e.,  scalar fields take the mean field $\langle \sigma \rangle$ and $\langle \xi \rangle$ which are constant over time and space, and for vector fields,  only  temporal component are considered $\langle \phi^0 \rangle$ and $\langle \omega^0 \rangle$ and the spatial components are assumed to be zero. For simplicity, the bracket $\langle \cdot \rangle$ and superscript are ignored, then the grand potential has form:
\begin{eqnarray}
    \Omega_{\text{Wal}} &=&\frac{1}{2}m_{\sigma}^2\sigma^2+\frac{1}{3}b_{\sigma}M_N(g_{\sigma}\sigma)^3+\frac{1}{4}c_{\sigma}(g_{\sigma}\sigma)^4\nonumber\\
    &-&\frac{1}{2}m_{\xi}^2\xi^2-\frac{1}{2}m_{\omega}^2\omega^2-\frac{1}{2}m_{\phi}^2\phi^2\nonumber\\
    &-&2T\sum_{B=N,H}\int \frac{d^3 k}{(2\pi)^3}\biggl[ \text{ln}\biggl( 1+e^{-\frac{ E^{*}_{B}(k)-\mu^{*}_B}{T}} \biggr ) \nonumber\\
    &&+\text{ln}\biggl( 1+e^{-\frac{ E^{*}_{B}(k)+\mu^{*}_B}{T}}\biggr )  \biggr],
\end{eqnarray}
and it's clear that the scalar fields contribute to baryons as effective mass: $E^{*}_{N/H}(k)=\sqrt{M_{N/H}^{* 2}+k^2}$, where the effective mass $M_{N/H}^{*}=M_{N/H}-g_{\sigma}\sigma-g_{\xi}\xi$, while vector fields serves as an extra baryon chemical potential: $\mu_B^{*}=\mu_B-g_{\omega}\omega-g_{\phi}\phi$. The mass of baryons take the vacuum value: $m_\sigma=550$MeV, $m_{\omega}=783$MeV, $m_{\phi}=1020$MeV, $m_{\xi}=975$MeV, $M_N=939$MeV, $M_H=1116$MeV.  Notice that nucleons are composed of $u$ and $d$ quarks while $s$ quark is also inside hyperons from quark model, it's reasonable that  $\sigma$ contributes to both nucleons and hyperons while $\xi$ only to the latter, and so are $\omega$ and $\phi$, thus, $g_{\xi N}=g_{\phi N}=0$, and the rest parameters can be found in Ref.\cite{Bunta:2004ej}. Similar to the NJL model, now we have a set of gap equations to solve:
\begin{equation}
    \frac{\partial\Omega_{\text{Wal}}}{\partial \sigma}=\frac{\partial \Omega_{\text{Wal}}}{\partial\xi}=\frac{\partial \Omega_{\text{Wal}}}{\partial \omega}=\frac{\partial \Omega_{\text{Wal}}}{\partial \phi}=0,
\end{equation}
after which the pressure can be obtained $P_B(T,\mu_B)=-\Omega_{\text{Wal}}(T,\mu_B)$. By Gibbs condition, the pressure at given temperature and baryon chemical potential is determined through $P=\text{Max}\{P_q,P_B\}$, and the $n$-th order susceptibilities of baryon number are defined  $\chi_n^B=\partial^n [p/T^4]/\partial [\mu_B/T]^n$, then the ratio of cumulants $C_4/C_2=\kappa\sigma^2=\chi_4/\chi_2$ is obtained.

\begin{figure}
 \centering
	\includegraphics[width=0.5\textwidth, trim=0 0 5 0, clip]{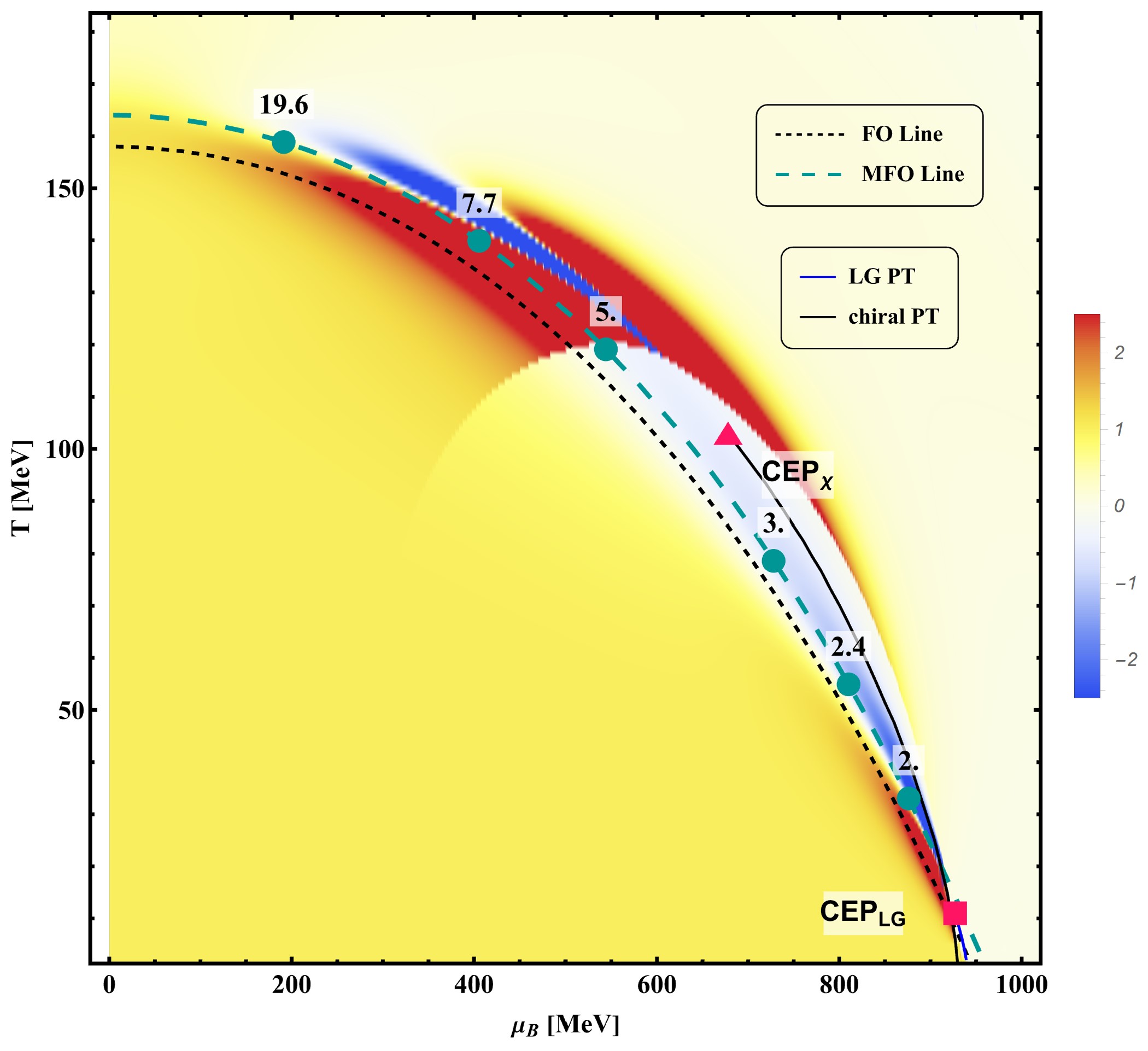}
	\begin{minipage}[c]{8cm}
	\caption{\justifying $(T,\mu_B)$ phase diagram from the hybrid model. $\kappa\sigma^2$ with magnitude in color and two different freeze-out lines, FO Line and MFO Line, in the $T-\mu_B$ plane. FO is fitted to experiment data\cite{Luo:2017faz} while MFO is modified by lattice, the expressions can be found in the context. The red triangle ($T_{\chi}^{\rm CEP}=103\text{MeV},\mu_{B,\chi}^{\rm CEP}=679\text{MeV}$) and square ($T_{LG}^{\rm CEP}=11\text{MeV},\mu_{B,LG}^{\rm CEP}=927\text{MeV}$) are CEPs for chiral and LG phase transition, respectively. The green dots denotes the $\sqrt{s}=19.6, 7.7, 5, 3, 2.4,2$GeV in MFO, and the transition of dominant degree between quarks and baryons occurs around at $\sqrt{s}\approx 5\text{GeV}$. And black and black solid lines are chiral and LG phase transition lines, respectively.}
\label{fig:densityplot}
\end{minipage}
\end{figure}

\begin{figure}
 \centering
	\includegraphics[width=0.5\textwidth, trim=0 20 0 20 clip]{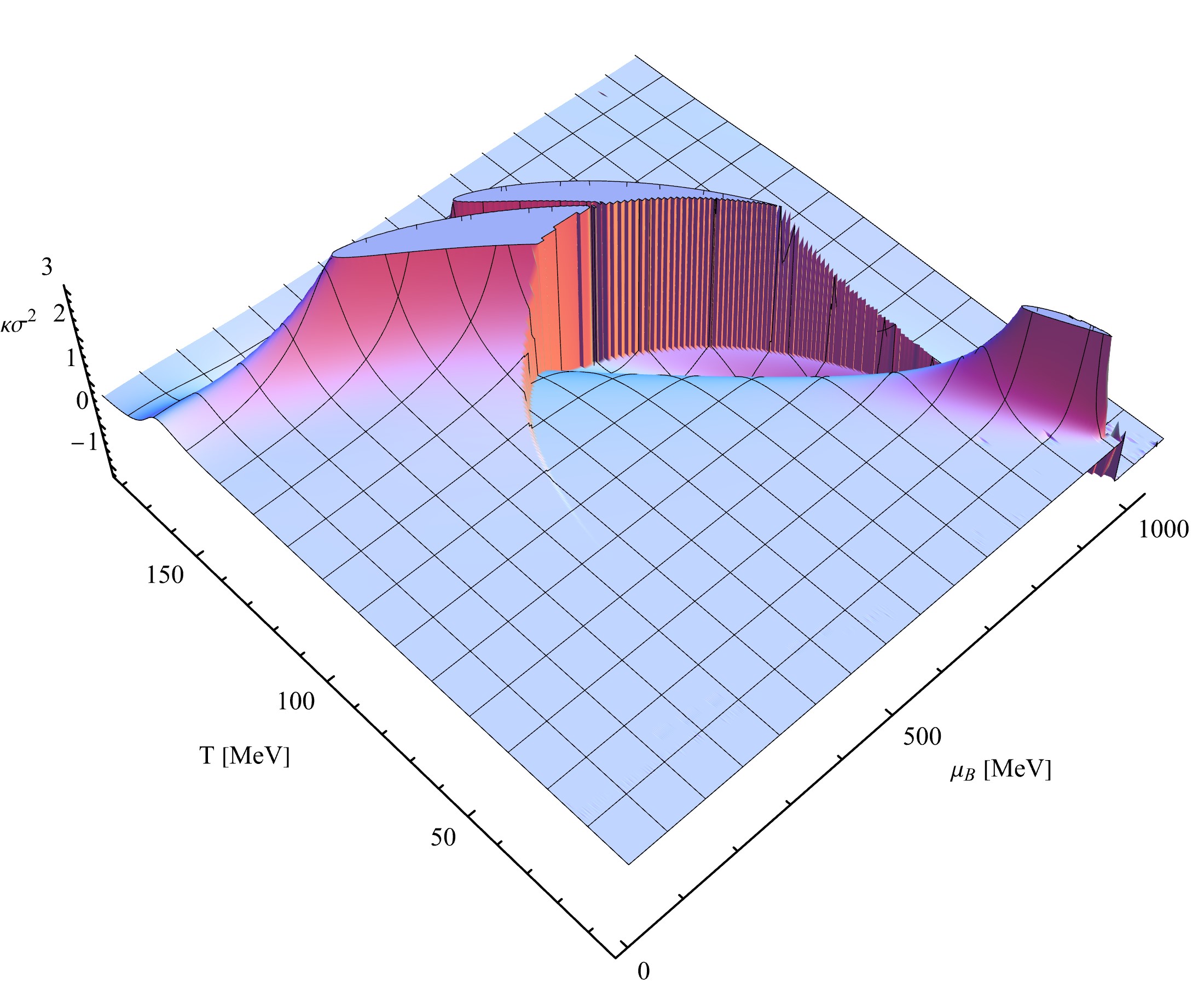}
	\caption{\justifying The 3D plot of  $\kappa\sigma^2$ in the hybrid model in $T-\mu_B$ plane.}
\label{fig:plot3D}
\end{figure}

\textit{Results:}  To see how the baryon number fluctuation varies with collision energy, we plot the $\kappa\sigma^2$ in the full $T-\mu_B$ plane as well as the freeze-out lines FO and MFO, as shown in Fig.\ref{fig:densityplot}. Numerical calculation shows that the PNJL model exhibits a CEP for chiral phase transition located at ($T_{\chi}^{\rm CEP}=103\text{MeV},\mu_{B,\chi}^{\rm CEP}=679\text{MeV}$), while the Walecka model exhibits another CEP at the end of LG phase transition ($T_{LG}^{\rm CEP}=11\text{MeV},\mu_{B,LG}^{\rm CEP}=927\text{MeV}$), which are denoted as red triangle and square in Fig.\ref{fig:densityplot}, respectively. The corresponding magnitude of $\kappa\sigma^2$ obtained in the hybrid model is marked in color in Fig.\ref {fig:densityplot}, the corresponding 3D plot of $\kappa\sigma^2$ from the hybrid model in the $(T,\mu_B)$ plane is shown in Fig.\ref{fig:plot3D}. It is noticed from Fig.\ref {fig:densityplot} and Fig.\ref{fig:plot3D} that the existence of LG phase transition as well as the $\text{CEP}_{LG}$ give a large region of negative $\kappa\sigma^2$. If the freeze-out line crosses the crossover region of LG phase transition, negative $\kappa\sigma^2$ can be observed in experiment.

%The hybrid model gives two CEPs, ($T_{\chi}^{\rm CEP}=103\text{MeV},\mu_{B,\chi}^{\rm CEP}=679\text{MeV}$) for chiral phase transition and ($T_{LG}^{\rm CEP}=11\text{MeV},\mu_{B,LG}^{\rm CEP}=927\text{MeV}$) for nuclear liquid-gas phase transition, which are shown in Fig.\ref {fig:densityplot} by red triangle and square, respectively. 

To make a direct comparison with experimental data, we calculated $\kappa \sigma^2$ along the freeze-out line, where the baryon chemical potential as a function of collision energy is obtained through experiments \cite{Luo:2017faz}(in GeV): $\mu_B = 1.477/(1+0.343\sqrt{s})$ as well as the baryon chemical potential and temperature: $T=0.158-0.14\mu_B^2-0.04\mu_B^4$ (FO). However, since we have used effective models that have discrepancies from the real world, and notice the relation of freeze-out line and the phase transition line (crossover) is crucial to the behavior of $\kappa\sigma^2$ as a function of collision energy \cite{Li:2018ygx}, it is reasonable to use freeze-out lines with slight deviations from the one fitted to experiments. At zero baryon chemical potential, the pseudo critical temperature of crossover is determined by Lattice QCD \cite{HotQCD:2018pds} with physical light and strange quark masses: $T_{pc}^{\text{LQCD}}=156.5\pm 1.5\text{MeV}$, and the above realistic PNJL model gives pseudo critical temperature as $T_{pc}^{{\rm PNJL}}=162.5\text{MeV}$. Thus, we define a new freeze-out line modified by LQCD, which has form of  $T=0.158-0.14\mu_B^2-0.04\mu_B^4+\Delta T$(MFO, Modified Freeze-Out line), with $\Delta  T=T_{pc}^{{\rm PNJL}}-T_{pc}^{\text{LQCD}}$. And the comparison between the theoretical results along MFO and experimental data would be realistic.

\begin{figure}
 \centering
	\includegraphics[width=0.5\textwidth]{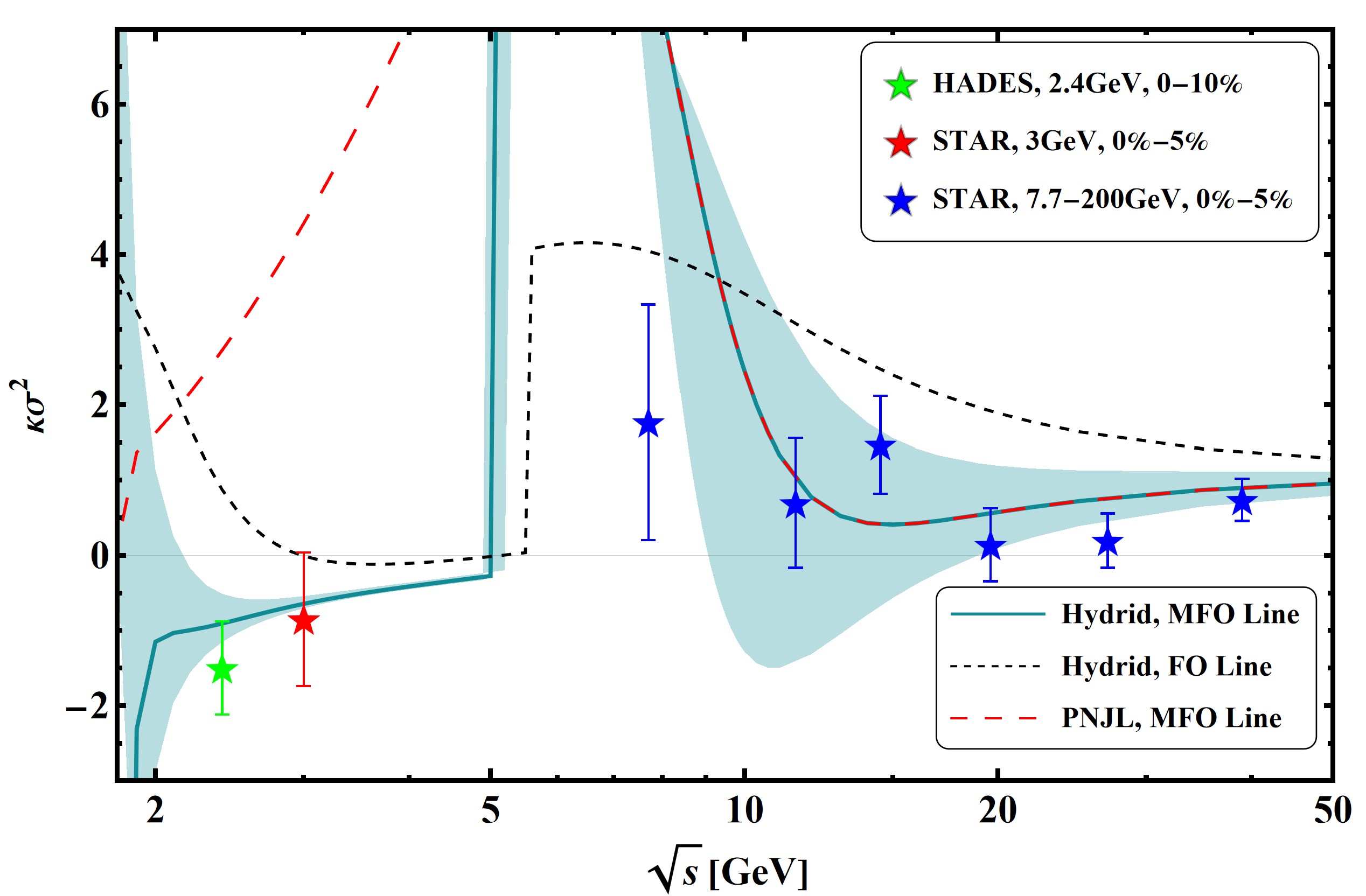}
	\caption{\justifying  The baryon number fluctuation $\kappa\sigma^2$ in the hybrid model as a function collision energy along two different freeze-out lines. And the green band is due to the $T_{pc}^{\text{LQCD}}$ uncertainty. The results of PNJL model is also plotted as the red dashed line shown. The stars are experiment data from STAR and HADES. 7.7GeV-200GeV of STAR rapidity $|y|<0.5$, 3GeV with $-0.5<y<0$, 2.4GeV of HADES $y=y_0\pm0.2$ where $y_0=0.74$ is Au + Au center-of-mass rapidity\cite{HADES:2020wpc}.}
\label{fig:kappasigma2}
\end{figure}
Now we focus on theoretical results along MFO, and the green dots show the locations of $\sqrt{s}=19.6, 7.7, 5, 3, 2.4, 2$GeV in the phase diagram along MFO. At collision energy $\sqrt{s}\gtrsim 5$GeV, quarks are the main dominant degree and $\kappa\sigma^2$ is affected mainly by the chiral phase transition: the slight decreases at $\sqrt{s}\gtrsim 10$ GeV is due to that MFO across the crossover region while the rapid increase  at $\sqrt{s}\sim 7$ GeV is a result of approaching to $\text{CEP}_{\chi}$. If the nuclear matter is not taken into account, then $\kappa\sigma^2$ will keep increasing until pass $\text{CEP}_{\chi}$. However, the hybrid model indicates that the dominant degrees of freedom change from quarks to baryons at $\sqrt{s}\sim 5$GeV shown in Fig.\ref{fig:densityplot}. At the crossover side of the the nuclear LG phase transition exhibits a region with negative value of $\kappa\sigma^2$, as the light and dark blue region shown in Fig.\ref{fig:densityplot}, thus, a jump of $\kappa\sigma^2$ from positive to negative occurs at $\sqrt{s}\sim 5$GeV, and continue to decrease at lower collision energy $\sqrt{s}\sim 2 {\rm GeV}$ as it approaches to $\text{CEP}_{\rm LG}$. For even lower energy, $\kappa\sigma^2$ could increase to around zero, which will be determined by the relative location of the critical baryon chemical potential $\mu_{B,\chi}^c(T=0)$ and $\mu_{B,{\rm LG}}^c(T=0)$ at zero temperature.

The results of $\kappa\sigma^2$ as a function of collision energy along two different freeze-out lines(FO and MFO) are shown in Fig.\ref{fig:kappasigma2}, and the 3.3GeV and 7.7-200GeV collision data of STAR, as well as the 2.4GeV collision from HADES are also shown, and only the most-central collisions are taken into consideration. It is seen that the FO shows a large deviation compared to experiment data, while the modified freeze-out line MFO captures the features of experiment data correctly and shows good agreement. As the collision energy decreases at the region $\sqrt{s}\gtrsim 5$GeV,  $\kappa\sigma^2$ starts from 1 and  decreases slightly and then increases dramatically at $\sqrt{s}\sim 10\text{GeV}$; at $\sqrt{s} \approx 5$GeV, $\kappa\sigma^2$ drops from positive value to negative and decreases with collision energy decreases in $2\text{GeV}\lesssim\sqrt{s}\lesssim 5\text{GeV}$. The result of the PNJL model or with only the chiral phase transition alone is also plotted, as the red dashed line in Fig.\ref{fig:kappasigma2}. For $\sqrt{s}\gtrsim 5$GeV, there is no difference between PNJL and Hybrid model， while at $\sqrt{s}\lesssim 5$GeV， the PNJL model still gives positive $\kappa\sigma^2$ which are opposite to experimental data. This indicates the importance of taking into account of the nuclear LG phase transition at high baryon density and low temperature.

\textit{Conclusion and Outlook:} We have investigated the baryon number fluctuation $\kappa\sigma^2$ as a function of collision energy in a hybrid model including both chiral phase transition and nuclear Liquid-Gas phase transition with corresponding ${\rm CEP}_{\chi}$ and ${\rm CEP}_{\rm LG}$ located at ($T_{\chi}^{\rm CEP}=103\text{MeV},\mu_{B,\chi}^{\rm CEP}=679\text{MeV}$) and ($T_{LG}^{\rm CEP}=11\text{MeV},\mu_{B,LG}^{\rm CEP}=927\text{MeV}$), respectively. The existence of LG phase transition led to a large region of negative $\kappa\sigma^2$ at large baryon chemical potential and low temperature, and the freeze-out line crosses this region, thus negative $\kappa\sigma^2$ can be observed in experiment. The numerical calculation shows that the baryon number fluctuation $\kappa\sigma^2$ along the freeze-out line are well in agreement with experimental data. The agreement between theoretical results and experimental data at collision energy $\sqrt{s}=3\text{GeV}$ and $2.4\text{GeV}$ indicates the existence of nuclear matter at high baryon density and near zero temperature as well as a first-order nuclear Liquid-Gas phase transition. 

As we can see that the baryon number fluctuation $\kappa\sigma^2$ can be used to probe the phase structure at high baryon density and low temperature, such as to confirm the possibility of an extra CEP proposed in \cite{Hatsuda:2006ps} with a crossover hadron-quark transition at zero temperature and high baryon density, which is important to determine the neutron star structure \cite{Huang:2022mqp} as well as the merging of neutron stars. Also the baryon number fluctuation $\kappa\sigma^2$ can be used to detect the quarkyonic phase as proposed in \cite{McLerran:2007qj,McLerran:2007qj}.

One can argue that in this work, the non-equilibrium evolution of the system \cite{Mukherjee:2015swa,Mukherjee:2016kyu} has not been discussed. At the moment, we can regard that the system at freeze-out is in equilibrium \cite{Braun-Munzinger:2003pwq}, therefore, whatever the evolution of the system, it always reaches equilibrium at freeze-out, therefore, the equilibrium description at freeze-out should be valid. On the other hand, we are connecting the microscopic model with hydrodynamic evolution in the ongoing project \cite{shen:2023abc}, 
where we will check how non-equilibrium evolution of the system will affect the observable at freeze-out.

\textit{Acknowledgements:} We thank Guoyun Shao for helpful discussion on Walecka model. This work is supported in part by the National Natural Science Foundation of China (NSFC) Grant Nos: 12235016, 12221005, 12147150 and the Strategic Priority Research Program of Chinese Academy of Sciences under Grant No XDB34030000, the start-up funding from University of Chinese Academy of Sciences(UCAS), and the Fundamental Research Funds for the Central Universities.

\bibliographystyle{unsrt}
\bibliography{reference.bib}

\begin{thebibliography}{10}

\bibitem{Pisarski:1983ms}
Robert~D. Pisarski and Frank Wilczek.
\newblock {Remarks on the Chiral Phase Transition in Chromodynamics}.
\newblock {\em Phys. Rev. D}, 29:338--341, 1984.

\bibitem{Stephanov:1998dy}
Misha~A. Stephanov, K.~Rajagopal, and Edward~V. Shuryak.
\newblock {Signatures of the tricritical point in QCD}.
\newblock {\em Phys. Rev. Lett.}, 81:4816--4819, 1998.

\bibitem{Hatta:2002sj}
Yoshitaka Hatta and Takashi Ikeda.
\newblock {Universality, the QCD critical / tricritical point and the quark
  number susceptibility}.
\newblock {\em Phys. Rev. D}, 67:014028, 2003.

\bibitem{Stephanov:1999zu}
Misha~A. Stephanov, K.~Rajagopal, and Edward~V. Shuryak.
\newblock {Event-by-event fluctuations in heavy ion collisions and the QCD
  critical point}.
\newblock {\em Phys. Rev. D}, 60:114028, 1999.

\bibitem{Hatta:2003wn}
Y.~Hatta and M.~A. Stephanov.
\newblock {Proton number fluctuation as a signal of the QCD critical endpoint}.
\newblock {\em Phys. Rev. Lett.}, 91:102003, 2003.
\newblock [Erratum: Phys.Rev.Lett. 91, 129901 (2003)].

\bibitem{Schwarz:1999dj}
T.~M. Schwarz, S.~P. Klevansky, and G.~Papp.
\newblock {The Phase diagram and bulk thermodynamical quantities in the NJL
  model at finite temperature and density}.
\newblock {\em Phys. Rev. C}, 60:055205, 1999.

\bibitem{Zhuang:2000ub}
P.~Zhuang, M.~Huang, and Z.~Yang.
\newblock {Density effect on hadronization of a quark plasma}.
\newblock {\em Phys. Rev. C}, 62:054901, 2000.

\bibitem{STAR:2010mib}
M.~M. Aggarwal et~al.
\newblock {Higher Moments of Net-proton Multiplicity Distributions at RHIC}.
\newblock {\em Phys. Rev. Lett.}, 105:022302, 2010.

\bibitem{STAR:2010vob}
M.~M. Aggarwal et~al.
\newblock {An Experimental Exploration of the QCD Phase Diagram: The Search for
  the Critical Point and the Onset of De-confinement}.
\newblock 7 2010.

\bibitem{STAR:2013gus}
L.~Adamczyk et~al.
\newblock {Energy Dependence of Moments of Net-proton Multiplicity
  Distributions at RHIC}.
\newblock {\em Phys. Rev. Lett.}, 112:032302, 2014.

\bibitem{Luo:2017faz}
Xiaofeng Luo and Nu~Xu.
\newblock {Search for the QCD Critical Point with Fluctuations of Conserved
  Quantities in Relativistic Heavy-Ion Collisions at RHIC : An Overview}.
\newblock {\em Nucl. Sci. Tech.}, 28(8):112, 2017.

\bibitem{STAR:2020tga}
J.~Adam et~al.
\newblock {Nonmonotonic Energy Dependence of Net-Proton Number Fluctuations}.
\newblock {\em Phys. Rev. Lett.}, 126(9):092301, 2021.

\bibitem{STAR:2022etb}
Mohamed Abdallah et~al.
\newblock {Higher-order cumulants and correlation functions of proton
  multiplicity distributions in sNN=3~GeV~Au+Au collisions at the RHIC STAR
  experiment}.
\newblock {\em Phys. Rev. C}, 107(2):024908, 2023.

\bibitem{Hatsuda:2006ps}
Tetsuo Hatsuda, Motoi Tachibana, Naoki Yamamoto, and Gordon Baym.
\newblock {New critical point induced by the axial anomaly in dense QCD}.
\newblock {\em Phys. Rev. Lett.}, 97:122001, 2006.

\bibitem{Huang:2022mqp}
Yong-Jia Huang, Luca Baiotti, Toru Kojo, Kentaro Takami, Hajime Sotani, Hajime
  Togashi, Tetsuo Hatsuda, Shigehiro Nagataki, and Yi-Zhong Fan.
\newblock {Merger and Postmerger of Binary Neutron Stars with a Quark-Hadron
  Crossover Equation of State}.
\newblock {\em Phys. Rev. Lett.}, 129(18):181101, 2022.

\bibitem{McLerran:2007qj}
Larry McLerran and Robert~D. Pisarski.
\newblock {Phases of cold, dense quarks at large N(c)}.
\newblock {\em Nucl. Phys. A}, 796:83--100, 2007.

\bibitem{Stephanov:2011pb}
M.~A. Stephanov.
\newblock {On the sign of kurtosis near the QCD critical point}.
\newblock {\em Phys. Rev. Lett.}, 107:052301, 2011.

\bibitem{Fu:2016tey}
Wei-jie Fu, Jan~M. Pawlowski, Fabian Rennecke, and Bernd-Jochen Schaefer.
\newblock {Baryon number fluctuations at finite temperature and density}.
\newblock {\em Phys. Rev. D}, 94(11):116020, 2016.

\bibitem{Fan:2016ovc}
Wenkai Fan, Xiaofeng Luo, and Hong-Shi Zong.
\newblock {Mapping the QCD phase diagram with susceptibilities of conserved
  charges within Nambu\textendash{}Jona-Lasinio model}.
\newblock {\em Int. J. Mod. Phys. A}, 32(11):1750061, 2017.

\bibitem{Li:2018ygx}
Zhibin Li, Kun Xu, Xinyang Wang, and Mei Huang.
\newblock {The kurtosis of net baryon number fluctuations from a realistic
  Polyakov\textendash{}Nambu\textendash{}Jona-Lasinio model along the
  experimental freeze-out line}.
\newblock {\em Eur. Phys. J. C}, 79(3):245, 2019.

\bibitem{HADES:2020wpc}
J.~Adamczewski-Musch et~al.
\newblock {Proton-number fluctuations in $\sqrt {s_{NN}}$ =2.4 GeV Au + Au
  collisions studied with the High-Acceptance DiElectron Spectrometer (HADES)}.
\newblock {\em Phys. Rev. C}, 102(2):024914, 2020.

\bibitem{Klevansky:1992qe}
S.~P. Klevansky.
\newblock {The Nambu-Jona-Lasinio model of quantum chromodynamics}.
\newblock {\em Rev. Mod. Phys.}, 64:649--708, 1992.

\bibitem{Vovchenko:2016rkn}
Volodymyr Vovchenko, Mark~I. Gorenstein, and Horst Stoecker.
\newblock {van der Waals Interactions in Hadron Resonance Gas: From Nuclear
  Matter to Lattice QCD}.
\newblock {\em Phys. Rev. Lett.}, 118(18):182301, 2017.

\bibitem{Glendenning:1992vb}
Norman~K. Glendenning.
\newblock {First order phase transitions with more than one conserved charge:
  Consequences for neutron stars}.
\newblock {\em Phys. Rev. D}, 46:1274--1287, 1992.

\bibitem{Shao:2017yzv}
Guo-yun Shao, Zhan-duo Tang, Xue-yan Gao, and Wei-bo He.
\newblock {Baryon number fluctuations and the phase structure in the PNJL
  model}.
\newblock {\em Eur. Phys. J. C}, 78(2):138, 2018.

\bibitem{He:2022yrk}
Wei-bo He, Guo-yun Shao, and Chong-long Xie.
\newblock {Speed of sound and liquid-gas phase transition in nuclear matter}.
\newblock {\em Phys. Rev. C}, 107(1):014903, 2023.

\bibitem{Bhattacharyya:2016jsn}
Abhijit Bhattacharyya, Sanjay~K. Ghosh, Soumitra Maity, Sibaji Raha, Rajarshi
  Ray, Kinkar Saha, and Sudipa Upadhaya.
\newblock {Reparametrizing the
  Polyakov\textendash{}Nambu\textendash{}Jona-Lasinio model}.
\newblock {\em Phys. Rev. D}, 95(5):054005, 2017.

\bibitem{Ghosh:2007wy}
Sanjay~K. Ghosh, Tamal~K. Mukherjee, Munshi~Golam Mustafa, and Rajarshi Ray.
\newblock {PNJL model with a Van der Monde term}.
\newblock {\em Phys. Rev. D}, 77:094024, 2008.

\bibitem{Bunta:2004ej}
Juraj~Kotulic Bunta and Stefan Gmuca.
\newblock {Hyperons in a relativistic mean-field approach to asymmetric nuclear
  matter}.
\newblock {\em Phys. Rev. C}, 70:054309, 2004.

\bibitem{HotQCD:2018pds}
A.~Bazavov et~al.
\newblock {Chiral crossover in QCD at zero and non-zero chemical potentials}.
\newblock {\em Phys. Lett. B}, 795:15--21, 2019.

\bibitem{Mukherjee:2015swa}
Swagato Mukherjee, Raju Venugopalan, and Yi~Yin.
\newblock {Real time evolution of non-Gaussian cumulants in the QCD critical
  regime}.
\newblock {\em Phys. Rev. C}, 92(3):034912, 2015.

\bibitem{Mukherjee:2016kyu}
Swagato Mukherjee, Raju Venugopalan, and Yi~Yin.
\newblock {Universal off-equilibrium scaling of critical cumulants in the QCD
  phase diagram}.
\newblock {\em Phys. Rev. Lett.}, 117(22):222301, 2016.

\bibitem{Braun-Munzinger:2003pwq}
Peter Braun-Munzinger, Krzysztof Redlich, and Johanna Stachel.
\newblock {Particle production in heavy ion collisions}.
\newblock pages 491--599, 4 2003.

\bibitem{shen:2023abc}
Yifan Shen, Wei Chen, Kun Xu, and Mei Huang.
\newblock {Hydrodynamical evolution for QCD equation of state with a critical
  end point}.

\end{thebibliography}

\end{document}